\documentclass[twocolumn, a4paper,10pt]{article}
\usepackage[top=2.5cm, bottom=2.5cm, left=2.0cm, right=2.0cm,
columnsep=0.8cm]{geometry}
\usepackage{enumitem}
\usepackage[hidelinks]{hyperref}
\usepackage{boxedminipage}
\usepackage{nopageno}
\usepackage{graphicx}
\usepackage[font=it]{caption}
\usepackage[usenames,dvipsnames]{xcolor}
\usepackage{listings}
\usepackage{caption}
\usepackage[utf8x]{inputenc} 
\usepackage{graphicx}      % include this line if your document contains figures
\usepackage{microtype}
\usepackage{graphicx}
\usepackage{booktabs} % for professional tables
\usepackage{amsfonts}
\usepackage{tikz}
\usepackage{changepage}
\usepackage{gensymb}
\usepackage{amsmath}
\usepackage{algorithm}
\usepackage{algorithmic}
\usepackage{pgf}
\usepackage{subfig}
\usepackage{varwidth}
\usepackage{longtable}
\newsavebox\tmpbox
\usetikzlibrary{arrows.meta, bending, backgrounds, scopes}
\usepackage{fancyhdr}

\setlist{itemsep=-0.1cm,topsep=0.1cm,labelsep=0.3cm}
\setenumerate{leftmargin=*}
\makeatletter
\renewcommand\title[1]{\gdef\@title{\fontsize{14pt}{2pt}\bfseries{#1}}}
\makeatletter
\renewcommand\section{\@startsection{section}{1}{\z@}{3pt}{3pt}{\normalfont\large\bfseries}}
\normalfont\large
\makeatletter
\renewcommand\subsection{\@startsection{subsection}{1}{\z@}{\z@}{\z@}{\normalfont\normalsize\bfseries}}
\makeatletter
\renewcommand\subsection{\@startsection{subsection}{1}{\z@}{\z@}{0.1pt}{\normalfont\normalsize\bfseries}}

\providecommand{\keywords}[1]
{
  \small	
  \textbf{\textit{Keywords---}} #1
}

\title{State space models for building control: how deep should you go?}

\author{
Baptiste Schubnel$^1$, Rafael E. Carrillo$^1$,  Paolo Taddeo$^2$, Lluc Canals Casals$^2$, Jaume Salom$^2$, \\
Yves Stauffer$^1$ and Pierre-Jean Alet$^1$\\ 																											% Line 4
$^1$Centre Suisse d'Electronique et de Microtechnique (CSEM), \\
 Neuchâtel, Switzerland\\ 																																	% Line 5
$^2$Catalonia Institute for Energy Research (IREC), \\
 Barcelona, Spain\\ 																									}								% Line 6

\date{\vspace{-0.5cm}}	% remove default date and replace the Blank 10th line
\begin{document}

\maketitle

\thispagestyle{fancy}
\chead{This is an Original Manuscript of an article published by Taylor \& Francis in\\Journal of Building Performance Simulation on 14 Sep 2020, available at \url{https://www.tandfonline.com/doi/10.1080/19401493.2020.1817149}}

\section*{Abstract}
%\addtocounter{section}{1}
Power consumption in buildings show non-linear behaviors that linear models cannot capture whereas recurrent neural networks (RNNs) can. This ability makes RNNs attractive alternatives for the model-predictive control (MPC) of buildings. However RNN models lack mathematical regularity which makes their use challenging in optimization problems. This work therefore systematically investigates whether using RNNs for building control provides net gains in an MPC framework. It compares the representation power and control performance of two architectures: a fully non-linear RNN architecture and a linear state-space model with non-linear regressor. The comparison covers five instances of each architecture over two months of simulated operation in identical conditions. The error on the one-hour forecast of temperature is  69\%  lower with the RNN model than with the linear one. In control the linear state-space model outperforms by 10\% on the objective function, shows 2.8 times higher average temperature violations, and needs a third of the computation time the RNN model requires. This work therefore demonstrates that in their current form RNNs do improve accuracy but on balance well-designed linear state-space models with non-linear regressors are best in most cases of MPC.
\vspace{2mm}

\keywords{Building modeling, Model Predictive Control, Linear state space models, Neural Networks, Optimization}

\setcounter{section}{0}
\section*{Introduction}

Following the success of deep learning for imaging, text, and audio processing,  architectures with (recurrent) neural networks are increasingly popular to model dynamic systems
 \cite{ogunmolu2016, gonzales2018,chung2014empirical}. Indeed they provide a gain in prediction accuracy over linear systems model that is evidenced by both theoretical experimental studies. For example, \cite{FUNAHASHI1993801} demonstrated universal approximation properties of recurrent neural networks (RNNs). Several groups have sought to take advantage of this gain by using RNNs as predictive models within model predictive control (MPC), an optimal control procedure for dynamic systems  \cite{camacho2007,peng2007,mayne2014,drgona2018,Colin,bieker2019}.

However these models are not guaranteed to be linear or convex. It makes it challenging to include them in optimisation problems. Convexity of the function encoded by an RNN can be enforced e.g., by imposing non-negative constraints on the weights and using rectified linear unit activations (ReLU). Such approaches show promising results for control applications \cite{amos2017input,chen2018optimal,chen2020input}. However these approaches come at the expense of the universal approximation capabilities of RNNs for non-convex system dynamics. Because of this trade-off, it is therefore an open question whether net gains in control performance can be obtained by using RNNs as predictors rather than linear state-space models.

Energy management in buildings is a representative illustration of this question. Indeed buildings are dynamic systems for which the common practice is to use rule-based controllers and the state of the art is MPC based on linear state-space models \cite{sturzenegger2016,Maarten2013}. These models are adequate and reliable for room temperature \cite{Tomasz,PEAN201935} but cannot capture some non-linear phenomena due to shading and solar gain, low-level control loops, or physical characteristics of technical systems such as heat pumps \cite{valenzuela2019}. The authors of \cite{Chen2017} used ``long short-term memory'' (LSTM), an RNN architecture, to model a complex simulated building and to minimize energy consumption through optimal control. They claimed some improvements over linear state-space models based on a limited amount of validation data. Therefore this paper sets about evaluating whether there are net, systematic gains in the performance of building control by using neural networks within MPC as compared to state-of-the-art linear state-space models.

In order to isolate the impact of control from random variations in e.g., weather and occupancy, and to run long experiments in identical conditions, we chose to apply the different control methods on a detailed physical model of a single building. We trained and used for control five instances for each of two types of models: (1) linear state-space models of the building envelope combined with a non-linear regressor to estimate energy consumption, and (2) fully non-linear, recurrent neural network models. In addition we implemented rule-based controllers as benchmarks. We evaluated the performance of the controllers over two months of operations.

Based on this experimental approach, the paper evaluates the relative strengths and weaknesses of linear state-space models and RNNs in terms of:
\begin{itemize}
\item Representation capabilities by training the different instances under several strategies and datasets
\item Control performance in a constrained optimization problem consisting in minimizing the power exchanged between the grid and a building with photovoltaic (PV) production under the constraint of thermal comfort in the building
\item Computing time.
\end{itemize}
None of the two approaches outperforms the other on all criteria. Our investigation confirmed the superior representation capability of RNNs. However, on balance, linear state-space models appear better suited to model-predictive control, except if the application emphasizes minimizing constraint violations.

The rest of the paper is organized as follows. Section \ref{sec1} describes the simulated building. Section \ref{sec2} presents the two types of models used in this study. Section \ref{sec3} explains the optimization strategy used to solve the non-convex problems within the MPC as well as the heuristics used to improve the performance of the objective minimization. Sections \ref{sec4} and \ref{sec5} respectively present the results of system identification and control performance results in Section \ref{sec5}. Section \ref{sec6} concludes the paper with a discussion and directions for future work. The appendix provides details on the hyperparameters of the models and the training parameters.

\section{Building test case}
\label{sec1}
A building simulated with EnergyPlus \cite{crawley2000energy} was used to carry out the systematic performance analysis. The simulation time resolution was equal to 3 minutes, but control variables were updated by the MPC every 15 minutes. The simulated building is a residential building with four apartments and eight thermal zones, located in Barcelona, Spain. It has two major technical systems: a centralized geothermal heat pump and a photovoltaic installation that has a peak power of 10.8kW.

\begin{figure}[H]
\begin{center}
\centerline{\includegraphics[scale=0.35]{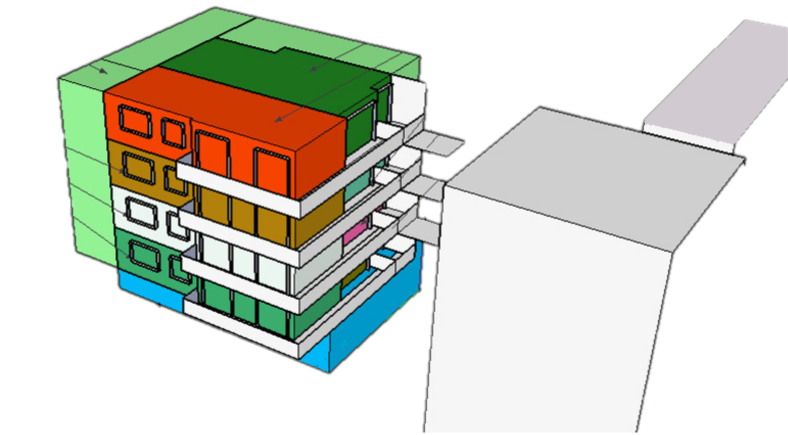}}
\caption{ \small  Building envelope overview with the four floors. There are two thermal zones per apartment.}
\label{bu}
\end{center}
\end{figure}
   
\normalsize
The centralized water-to-water geothermal heat pump system, which extracts heat from the ground through a vertical ground heat exchanger, provides hot water for the indoor fan coil units (two units per apartment) and the domestic hot water system. The domestic hot water system is composed of four storage tanks, one for each household. The heating loop circuit from the heat pump is connected to the bottom half part of the tanks and electrical heaters are placed on the top part acting as auxiliary heating systems to meet domestic hot water demand. The heat pump is connected to four tanks for hot water consumption and  to a heating loop for rooms thermal regulation. 

Room temperature control is carried out with thermostats and by a low-level control loop with an EnergyPlus Program Manager that implements an hysteresis control based on thermostats setpoints (SP) and bottom tank temperature setpoints. It triggers the on/off cycles of the heat pump based on setpoints hysteresis violations. The low-level control loop coupled with the heat pump is responsible for a non-linear behavior of the power consumption as a function of the room  thermostat temperatures and the pump supply temperature. The heat pump properties have been carefully calibrated to match the behavior of a real geothermal heat pump in the IREC laboratory. See \cite{taddeo2020management} for further details of the EnergyPlus model where its thermal envelope was used with a real heat pump in the IREC laboratory.

Daily schedule patterns for occupancy, hot water, light and appliances consumption are used to model activity in the building. Control variables are the room thermostats (4 setpoints, one per apartment), the tank bottom temperature (4 setpoints) and the heat pump supply temperature. Restrictions are imposed on the range of allowed values as displayed on Table \ref{table_sp}. The setpoint range for air temperatures is [19,24] \degree C to ensure occupants' thermal comfort. 
Moreover, the bottom temperatures of the tanks have to stay 5 \degree C  lower than the heat pump supply temperature.

\begin{table}[h]
\caption{ \small  Setpoints ranges}
\begin{center}
\label{table_sp}
\begin{small}
\begin{sc}
\begin{tabular}{lll}
\toprule
 \textbf{Controlled setpoints}     &  \textbf{Range (\degree C)}  \\
\midrule
Heat pump supply temp.  &   $[40,55]$\\ 
Tanks bottom temp. (4 SP)& $[35,50]$\\ 
Zone temp. (4 SP)& $[19,25]$\\ 
\bottomrule             
\end{tabular}
\end{sc}
\end{small}
\end{center}
\end{table}

\normalsize 
The photovoltaic (PV) installation is simulated as an array of PV modules with inverter. The PV array has an active surface area of $58 \rm{m}^2$ and is oriented South with an inclination of 40\degree. The full geometric model for solar radiation is used, including shading and reflections, to determine the incident solar resource. The EnergyPlus PV model was parameterized to match the real PV system installed on the IREC laboratory roof in Tarragona \cite{taddeo2020management}.

\section{Linear and non-linear state space models}
\label{sec2}
Two different model architectures were used to estimate rooms temperatures and electrical energy consumption: 
\begin{itemize}
\item The first model is a linear state-space model that is used to estimate room temperatures (see e.g., \cite{van2012subspace}). A non-linear regressor is used to estimate electrical energy consumption by the thermal system using state space model predictions. Model instances built with this architecture are called LSS-NL in the rest of the paper. 

\item The second model is a fully non-linear state-space model based on an encoder-decoder architecture with LSTMs \cite{lstm_art}. It  simultaneously estimates temperature in the rooms and electrical energy consumption by the thermal systems. Model instances built with this architecture are called ENC-DEC in the rest of the paper. 
\end{itemize}

\subsection{Linear state-space models with non-linear regression}
A  linear state-space model follows the  set of equations
\begin{equation}
\label{dyn_state_lin}
\begin{split}
x_{k+1} &= A x_{k} + B u_{k} + \varepsilon_k\\
y_{k} &= Cx_k + Du_{k} + \varepsilon'_k, 
\end{split}
\end{equation}
where $k$ is the discrete timestep, $x_{k} \in \mathbb{R}^{h}$ is the state variable at time $k$, $u_k \in \mathbb{R}^{ex}$ are the inputs at time $k$ (or exogenous variables),  $y_k \in \mathbb{R}^{end}$ are the modeled system outputs (endogenous variables) at time $k$ and $\varepsilon_k,  \varepsilon'_k$ are white random vectors with zero mean. Initially, internal state variables $x_k$ are estimated from a sequence of observations $y_1,...,y_n$ using the forward innovation Kalman equation \cite{van2012subspace}. For the building test case, $u_k$  in Eq. \eqref{dyn_state_lin} is a concatenation of the control commands and the external disturbances at time step $k$. The control commands are the temperature setpoints for the zone's thermostats, tanks and  heat pump forward temperature. The external disturbances are the solar irradiance, outdoor temperature and humidity.  The output variables $y_k$ in Eq. \eqref{dyn_state_lin} are the rooms temperatures. Even if the state variable $x_k$ in Eq. \eqref{dyn_state_lin} has a priori no physical meaning, correspondence between RC models and linear state space models for buildings leads to the interpretation of $x_k$ as representing effective dynamics of internal components e.g., walls and sub-layer temperatures.

Electrical energy consumption in buildings may exhibit non-linear behavior, in particular if systems like heat pumps and HVAC are used for heating and/or cooling the building. The non-linear behavior may also arise from low-level controllers that cannot be directly controlled by the MPC controller.  A kernel regression with radial basis functions was chosen to model the electric energy consumption. More specifically, at any time $k$, non-linear variables $z_k \in \mathbb{R}^{nl}$ (here the electric energy consumption) are modeled by 
\begin{equation}
\label{nl}
z_k = \sum_{i=1}^{N} \alpha_i \varphi(w_i,w_k)
\end{equation}
where $\varphi(x,y) = \exp(-\gamma \| y-x \|^2)$ is the kernel function, $w_k := (u_k,y_k)$ is the concatenation of the inputs and output variables, and $N$ is the number of points in the training set.  The right-hand side in Eq. \eqref{nl} is a closed-form expression which describes a smooth function. These two properties are advantages of  kernel regression over other non-linear regression methods ( see e.g. \cite{Debaditya2019,Sokratis2017}) if the optimization techniques used for the model predictive controller require gradient or hessian estimates of $f$, as discussed below in Section \ref{sec3}. 

\subsection{Non-linear state-space model}
The non-linear state space model is based on an encoder-decoder architecture where both the encoder and the decoder are  long short-term memory (LSTM) cells \cite{lstm_art}. The encoder is an LSTM-network that is used instead of a Kalman filter to initialize the state of the physical system. The decoder is another LSTM network followed by a multilayer perceptron (MLP) that  is used to predict the model outputs step by step.  Let $n \in \mathbb{N}$ be the number of encoder steps used to initialize the model.  A schematic view of the  neural architecture is given on Figure \ref{nnfig}.

\def\layersep{1.5cm}
\begin{figure}[ht]
\begin{adjustwidth*}{}{0em} 
    \begin{tikzpicture}[shorten >=1pt,->,draw=black!50, 
    node distance = \layersep,
every pin edge/.style = {<-,shorten <=1pt},
        neuron/.style = {circle,fill=black!25,minimum size=17pt,inner sep=0pt},
  input neuron/.style = {neuron, fill=green!50},
 output neuron/.style = {neuron, fill=red!50},
 hidden neuron/.style = {neuron, fill=blue!50},
         annot/.style = {text width=4em, text centered},
                         ]

 \node[input neuron, pin=left: $\tilde{w}_{t_{0}-n}$] (I-1) at (-2,-1) {};
 \node[input neuron, pin=left: $\tilde{w}_{t_{0}-n+1}$] (I-2) at (-2,-2) {};
 \node[input neuron, pin=left: $\tilde{w}_{t_{0}-1}$] (I-4) at (-2,-4) {};
 \node[input neuron, pin=left:  $ \ \vdots \ $ ] (I-3) at (-2,-3) {};

% Draw the decoder layer nodes
   \path[yshift=0.5cm]
    node[hidden neuron, pin=left: $u_{t_{0}}$] (H-5) at (-0.5,-5-0.5) {};
   \path[yshift=0.5cm]
    node[hidden neuron, pin=left: $u_{t_{0}+1}$] (H-6) at (-0.5,-6-0.5) {};
   \path[yshift=0.5cm]
    node[hidden neuron, pin=left: $ \ \vdots \  $] (H-7) at (-0.5,-7-0.5) {};
   \path[yshift=0.5cm]
    node[hidden neuron, pin=left: $u_{t'}$] (H-8) at (-0.5,-8-0.5) {};

\node[output neuron,pin={[pin edge={->}]right: $y(t_{0}),z(t_0)$}, right of=H-5] (O-5) {};
\node[output neuron,pin={[pin edge={->}]right: $y(t_{0}+1),z(t_0+1)$}, right of=H-6] (O-6) {};
\node[output neuron,pin={[pin edge={->}]right: $  \vdots \  $ }, right of=H-7] (O-7) {};
\node[output neuron,pin={[pin edge={->}]right: $y(t'),z(t')$}, right of=H-8] (O-8) {};

%Connect every node in the input layer with every node in the hidden layer.
 \path (I-1) edge (I-2);
 \path (I-2) edge (I-3);
 \path (I-3) edge (I-4);

 \path (H-5) edge (H-6);
 \path (H-6) edge (H-7);
 \path (H-7) edge (H-8);

% Connect every node in the decoder layer with Fc layer
\foreach \source in {5,...,8}
       \path (H-\source) edge (O-\source);

% Connect every node in the hidden layer with the output layer
%\foreach \source in {1,...,5}
 %   \path (H-\source) edge (O);

% Annotate the layers
 \node[draw] at (-2, -0.1)   (enc) {Encoder};
 \node[draw] at (-0.5, -4.1)   (dec) {Decoder};
 \node[draw] at (1.3, -4.1)   (fc) {MLP};

% curve from encoder to decoder
\scoped[on background layer]
\draw[line width=1mm, gray!50, shorten >=1mm, shorten <=1mm, -{Latex[flex]}]  (I-4) --  (H-5);
    \end{tikzpicture}

\end{adjustwidth*}
\caption{ \small  Structure of a non-linear state-space model based on an encoder-decoder architecture}
\label{nnfig}
\end{figure}
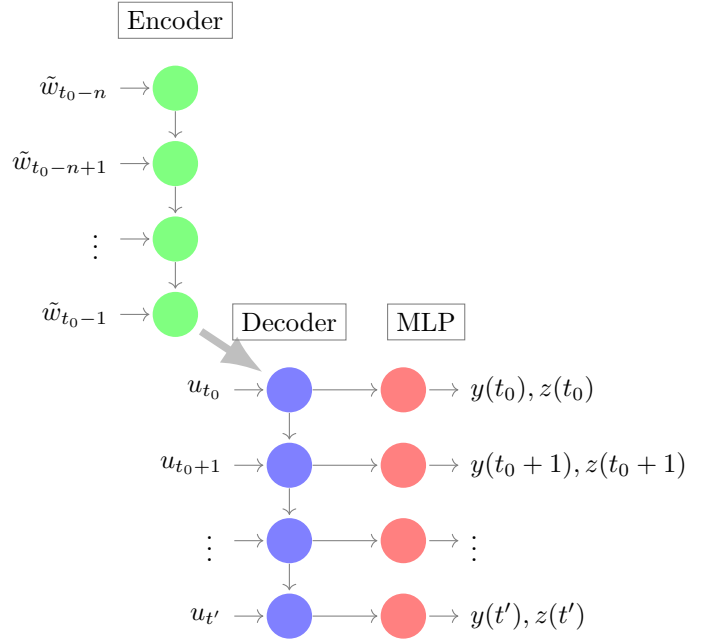

\normalsize
The encoder cell state and output at time $t$ are vectors of size $p$ respectively denoted by $c_{\tiny enc}(t) \in \mathbb{R}^{p}$ and $h_{\tiny enc}(t) \in \mathbb{ R}^{p}$.  The concatenation  $(c_{\tiny enc}, h_{\tiny enc}) $ plays a similar role as the state variable $x$ in linear state space models; see Eq. \eqref{dyn_state_lin}. Like the linear Kalman filter, the LSTM encoder is used to generate the representation of the initial system state at time $t_0$, given the past $n$ values. Setting $\tilde w := (u,y,z)$ as a concatenation of  exogenous variables,  room temperature variables and  energy variables, the encoder can be viewed as an iterative application of the map  $f_{\text{\tiny enc}} : \mathbb{R}^{ex + end+ nl+2p} \rightarrow \mathbb{R}^{2p}$ such that the tuple $(c_{\tiny enc}(t),h_{\tiny enc}(t))$ is a representation of the initial state of the system for $t=t_0-1$ provided that Eq. \ref{rec} has been repeated $n$ times from $t_0-n$ to $t_0-1$.
\begin{equation}
\label{rec}
(c_{\tiny enc}(t),h_{\tiny enc}(t)) = f_{\tiny enc}(\tilde{w}(t),c_{\tiny enc}(t-1),h_{\tiny enc}(t-1))
\end{equation}
For $t' \geq t_0$, the decoder LSTM is a map $ f_{\tiny dec} : \mathbb{R}^{ex+2p} \rightarrow \mathbb{R}^{2p}$ that only depends on the commands and external parameters (but not on previous system observables $y$):
\begin{equation}
\label{rec2}
(c_{\tiny dec}(t'),h_{\tiny dec}(t')) = f_{\tiny dec}(u(t'),c_{\tiny dec}(t'-1),h_{\tiny dec}(t'-1))
\end{equation}
where $c_{\tiny dec}(t_0-1) = c_{\tiny enc}(t_0-1)$ and  $h_{\tiny dec}(t_0-1) = h_{\tiny enc}(t_0-1)$. 

The output $h_{D}(t')$ is  fed to a multilayer perceptron network, giving rise to the estimated output $y(t')$ at time $t'$:
\begin{equation}
y(t'),z(t') = f_{MLP}(h_{D}(t')).
\end{equation}

Both architectures were trained by minimizing the mean square error between observations and predictions. For both architectures, the electric power production from photovoltaic panels was estimated independently using a simple physical model from the PVLIB library \cite{stein2016pvlib} since the geometry of the PV system at the IREC site was simple. The physical model has as input solar irradiance forecasts.

\subsection{Training method}
\label{subsecTrainingMethod}

The architectures have different sampling efficiency for system identification and different datasets were therefore used to identify the building models.  

Linear state space models identification techniques are  well covered in the  literature; see e.g., \cite{ljung2001system, van2012subspace}. Linear state space system identification is very sample efficient, theoretically founded,  and a few days of observations are sufficient for rooms temperature identification if signal excitations can be sent to the building. In this paper, the N4SID method is used to carry out rooms' dynamics identification; see \cite{van2012subspace}. Non-linear variables identification with kernel ridge regression is sample-efficient as well. 

On the other hand, the encoder-decoder architecture is sample inefficient and tends to be overparameterized (see e.g., \cite{Nakkiran2020Deep} for an interesting empirical study of over-parameterization in Neural Networks architecture and its effect on model accuracy). It needs to ingest data under various conditions and setpoints to avoid model overfit, and a period of at least a year of data is needed to perform good identification (to cover the four seasons). It is trained by stochastic gradient descent, with no optimal fitting guarantee because of the non-convexity of the architecture. 

For the linear state space model with non-linear outputs, between 7 and 40 days of data with  random multi-sine signals excitation ( see e.g., \cite{schoukens2019nonlinear}) were used to capture room dynamics and  electrical power from thermal sources. The distinct numbers of days considered as well as the random character of the excitations were used to build five models with distinct dynamic matrices $A,B, C$ and $D$. The models were kept constant over the two months evaluation phase, as degradation in models predictions were observed in case where refit of the matrices $A,B,C,D$ was carried out online with collected data from the MPC. The encoder-decoder model instances were trained on seven years of data with piecewise constant setpoints updated at random frequency and with random amplitudes lying in an acceptable physical range. Prediction range were arbitrarily sampled between one hour to one day ahead during training, to ensure that the model was able to capture short- and long-range dynamics. As for the linear state space models, five different models were built using stochasticity of the training by mini-batch gradient descent, as well as by varying the number of training steps. Models hyperparameters are detailed for all models in \ref{AA}.

\section{Formulation of the model predictive controller}
\label{sec3}
The objective of the model predictive controller is to minimize the power exchanged between the grid and the building without violating room temperature comfort constraints. The problem horizon is denoted by $H$ and the time resolution by $\Delta$.  Splitting the exogenous variables $u \equiv(u_{c},u_{nc})$ into controllable (setpoint inputs) and non-controllable variables (exterior parameters) and introducing the \emph{model function} $m: \mathbb{R}^{ex \times H} \rightarrow  \mathbb{R}^{(end +nl) \times H}$ that predicts room temperatures and non-linear variables (electrical energy consumption)  the optimization problem at every discrete optimization starting time $t$ takes the form
\begin{equation}
\label{optim_prob}
\begin{split}
&\underset{ \underline{u}_{c} \in \mathbb{R}^{ex_c \times H}}  {\min} \quad f(m( \underline{u}))\\
& s.t. \qquad g(m(\underline{u})) \leq 0, 
\end{split}
\end{equation}
where $f$ denotes the objective function and $g$ the constraint vector. The initialization phase is neglected here to simplify notations.

In the present case, the objective function $f$ is the mean absolute difference between the  electric power consumed by all equipment, $P_{\rm el}(t)$, and the PV power produced, $P_{ \rm prod}(t)$ i.e., 
\begin{equation}
f(m( \underline{u})) = \sum_{i=1}^{H} \vert P_{\rm el}(t+ i\Delta) -P_{ \rm  prod}(t + i\Delta) \vert,
\end{equation}
for $\underline{u} = (u(t+\Delta),..., u(t+H\Delta))$. The equipment comprises appliances, lighting system and thermal systems i.e., heat pump, auxiliary pumps and associated fans.
The components of the constraint vector $g$ are the room constraints $T_r-24\degree C$ and  $19\degree C -T_r$, that apply to all eight rooms, as well as the  setpoint constraints in Table \ref{table_sp}. As commonly done in MPC, problem \eqref{optim_prob} is solved at discrete optimization time $t$ over the entire horizon $H$ but only the first setpoint $u(t+\Delta)$ is applied, and a new problem is formulated starting at time $t+\Delta$ in a receding horizon fashion.

The problem in Eq. \eqref{optim_prob} is non-convex for both models and was solved using sequential quadratic programming (SQP)\cite{gill2012sequential}. SQP involves the minimization of a quadratic approximation of the objective function $f(m( \underline{u}))$ under a linear approximation of the constraint vector $g(m( \underline{u}))$. It solves iteratively the problem starting from an initial trajectory $\underline{u}^{(0)}$, making local quadratic expansion around each solution $\underline{u}^{(k)}$, $k\geq 0$,  and solving the quadratic  sub-problem \ref{quad_form} to obtain the next value $\underline{u}^{(k+1)}$. Iterations stop when the difference between two subsequent values becomes negligible or when a maximum number of iterations (e.g., 12) is reached.
\begin{equation}
\label{quad_form}
\begin{split}
&\underset{ \underline{u}_{c} \in \mathbb{R}^{ex_c \times H} } {\min} \quad f(m(\underline{u}^{(k)})) + \nabla (f \circ m)^{T}_{ \mid \underline{u}^{(k)}} (\underline{u}_c-\underline{u}^{(k)})  \\
& \qquad  \qquad + \frac{ 1}{2} (\underline{u}_c-\underline{u}^{(k)})^{T} Q(\underline{u}^{(k)})  (\underline{u}_c-\underline{u}^{(k)})\\
& s.t. \qquad g(m(\underline{u}^{(k)})) + \nabla (g \circ m)^{T}_{ \mid \underline{u}^{(k)}} (\underline{u}_c-\underline{u}^{(k)}) \leq 0, 
\end{split}
\end{equation}
for all $k \geq 0$. The matrix $Q$ denotes the Hessian of the objective $f \circ m$ of the original problem. In practical algorithms, important speed up and accuracy gains arise if the gradients of the objective and constraint functions can be computed without using finite difference methods. In the present work,  for the linear state space method with kernel regressor, the explicit  expression of the non-linear output is a net advantage in terms of speed and numerical stability as it can be differentiated easily - see Section \ref{perf} and \ref{excect}. The non-linear state space model can use the automatic differentiation capabilities of the deep learning libraries like Tensorflow or Pytorch \cite{tensorflow2015} to compute the model Jacobians. In that case, estimating gradients with Graphics Processing Unit (GPU) provides better and faster estimates than finite difference methods. Hessians are estimated using the LBFGS secant method \cite{fletcher1987}. 

Since the problem in Eq. \eqref{optim_prob} is non-convex, a key point to get good solutions with SQP is the choice of the initial solution trajectory $\underline{u}^{(0)}$. To stabilize the optimization process with the fully non-linear state-space model we used a shifted and smoothed version of the solution at the previous optimization step as initial trajectory. The resulting trajectories are presented in Section \ref{perf}.

\section{Identification results}
\label{sec4}
\begin{figure}[h]
  \centering
  \subfloat[ sMRAE from system identification \label{sMRAE}]{\includegraphics[width=0.8\columnwidth]{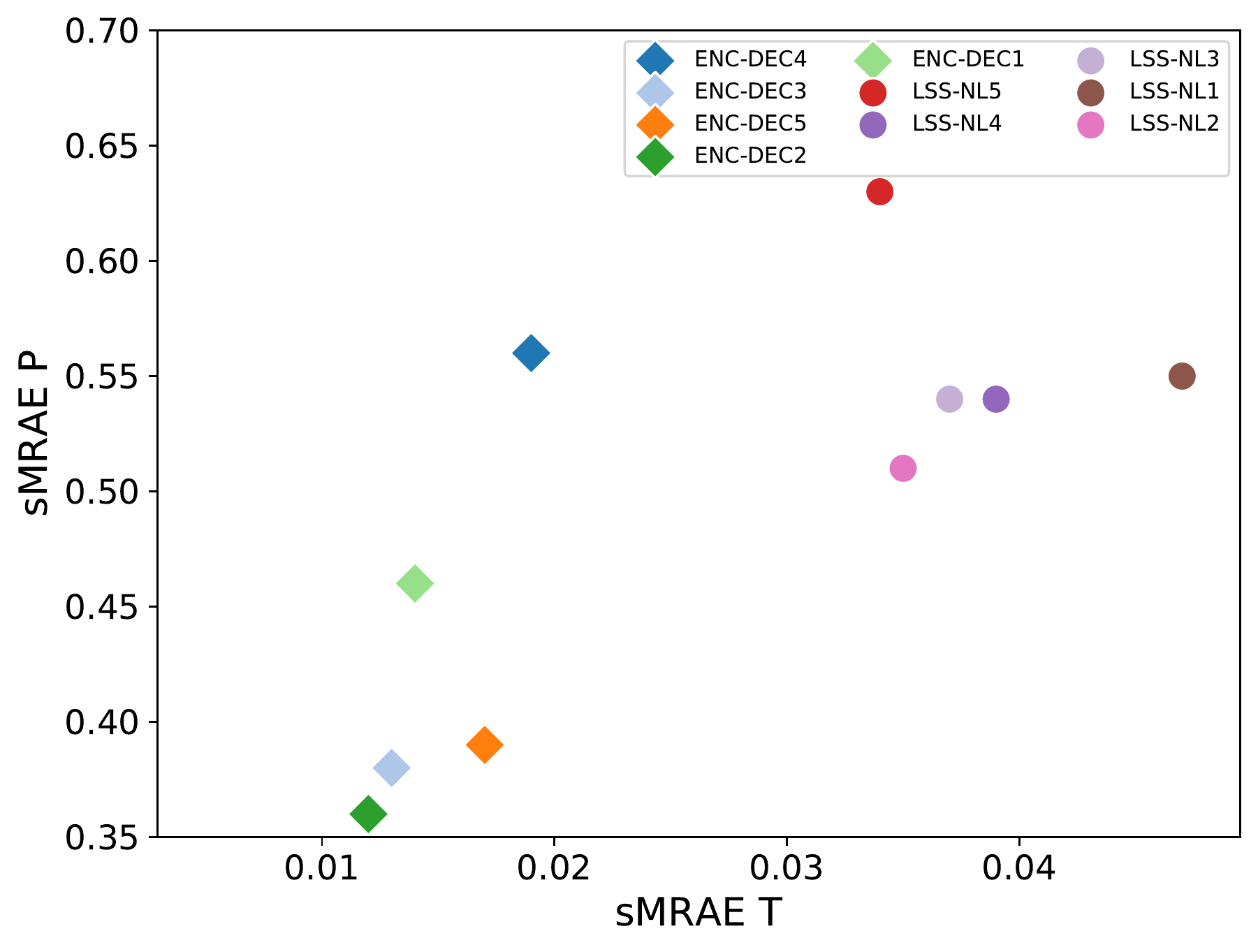}}
  \hfill
  \subfloat[Temperature error distribution \label{tmp_err}]{\includegraphics[width=0.8\columnwidth]{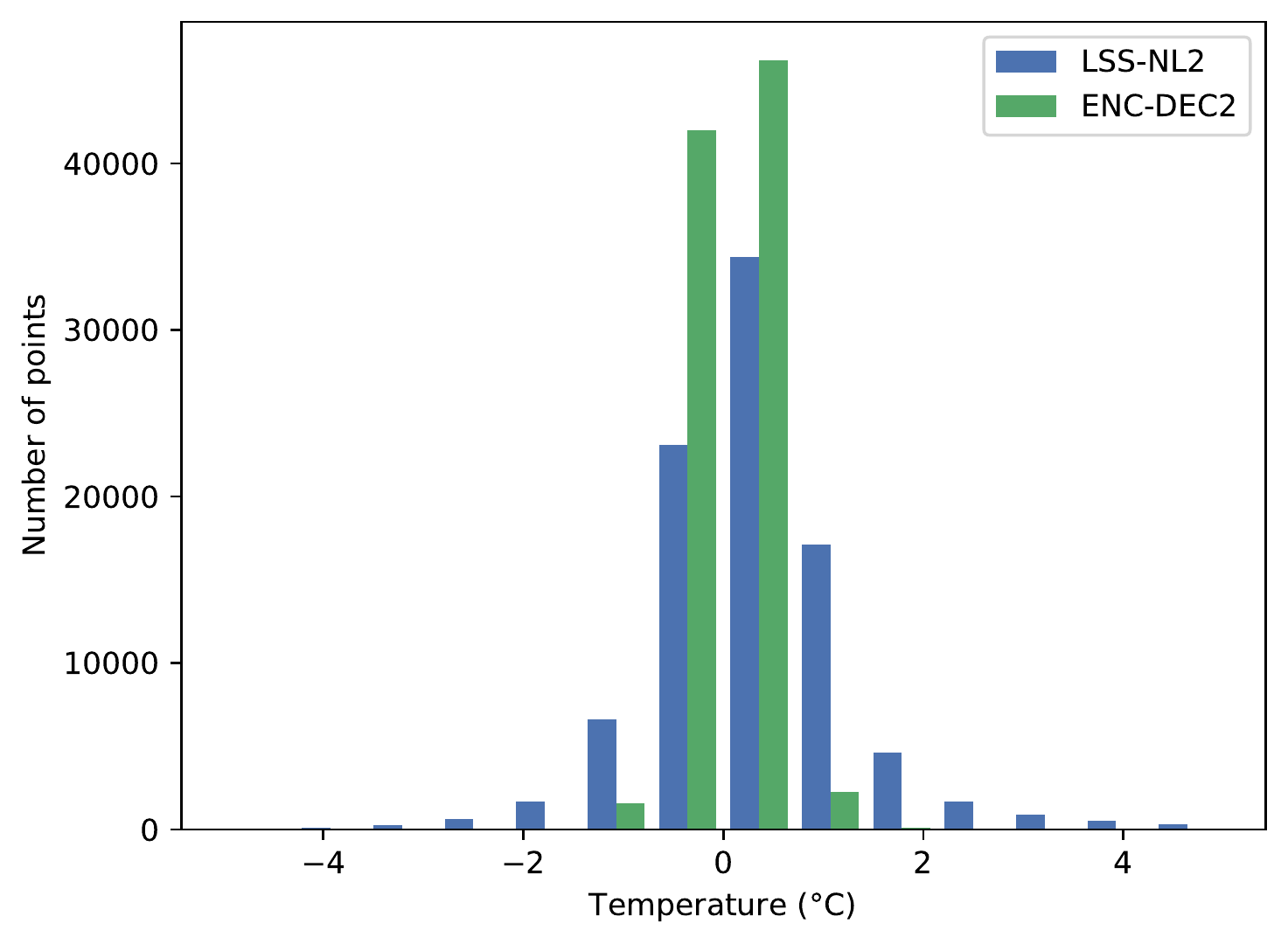}}
  \caption{ \small (a) sMRAE  for temperature and power predictions at one day ahead, for the five instances of each model architectures. ENC-DEC denotes encoder-decoder model instances, and LSS-NL linear state space model instances with non-linear power prediction; see Section \ref{sec2}. (b) Temperature error distribution for the best two model instances of each architecture at one day ahead.}
\end{figure}

To assess the representation capabilities of each architecture, models were evaluated on a test set of setpoints and external parameters not seen during the identification phase: a random profile of sinusoidal, square and triangular setpoints, with varying frequencies over time were created. The models were evaluated on four months of  data, from January to end of April in Barcelona. The evaluation windows were shifted on a fixed basis and prediction accuracy over short and long range were computed. At time t,  both LSS-NL and encoder-decoder models were used to predict  $m$ time steps ahead until $t+m$ given the last observed data points. Then the windows were shifted $m$ steps ahead and evaluation was performed starting at $t+m$, and so on. Results are displayed for $m=4$ (one-hour-ahead prediction) and $m=96$ (one-day-ahead prediction) on Figures \ref{sMRAE} and \ref{tmp_err}. In \ref{error_app}, all the numerical values of the mean absolute error (MAE) and symmetric mean relative absolute error (sMRAE) are tabulated. They are defined for $N$ predictions $\hat{y}(t_i)$ and observations $y(t_i)$ by
\begin{align}
\text{MAE} &= \frac{1}{N} \sum_{k=1}^{N} \vert \hat{y}(t_i) -y(t_i) \vert, \\
\text{sMRAE} &= \frac{2}{N} \sum_{k=1}^{N} \frac{\vert \hat{y}(t_i) -y(t_i) \vert}{\vert \hat{y}(t_i)\vert + \vert y(t_i) \vert}.
\end{align}

Encoder-decoder architectures (the five models labeled by ENC-DEC1,\dots, ENC-DEC5) exhibit significantly lower prediction errors both on one-hour and one-day-ahead predictions. For one-hour-ahead forecast, encoder-decoder architectures have a mean absolute error between 0.19 and 0.24 \degree C for temperature predictions and between 1.90 kW and 2.50 kW for power predictions, whereas linear state space models (LSS-NL1 to 5) have errors on temperature and power between 0.64 and 0.77 \degree C and between 3.24 kW and 4.32 kW, respectively (see Table \ref{tableOneHourErrorMetrics} i the Appendix). For comparison, the peak power generation by the PV system is 10.8 kW and the average power consumption under rule-based control is between 5 kW and 6 kW. For-day-ahead forecasts, errors grow but exhibit similar ratios between encoder-decoder models and linear state-space models (see Table \ref{tableOneDayErrorMetrics} in the Appendix). Figure \ref{tmp_err} shows the temperature error distributions for one-day-ahead predictions for the best model fits of both architectures. Power error distributions show  similar patterns, with the error distribution being broader for the linear state space models than the encoder-decoder architectures.

\normalsize

\section{MPC results}
\label{sec5}
\subsection{Control performance}
\label{perf}
Four metrics are used to estimate the control performance of the model predictive controllers. The two first metrics are directly related to the optimization objective and the constraints. The first metric is the  mean power exchanged with the grid, given by 
\begin{equation}
\overline{P}  = \frac{1}{N} \sum_{n=1}^{N} \vert P_{\rm el}(t_n) -P_{prod}(t_n) \vert.
\end{equation}
The second metrics is used to evaluate mean comfort bounds violation, and is defined by
\begin{equation}
\begin{split}
\overline{C} &:= \frac{1}{N} \sum_{n=1}^{N}  \sum_{r = 1}^{8}(\chi(T_r(t_n)>24) (T_r(t_n)-24) \\
& \text{}+ \chi(T_r(t_n)<19) (19-T_r(t_n)) ),
\end{split}
\end{equation}
where $T_r$ is the measured temperature of room $r$ and $\chi$ is the Heaviside step function.
The metric $\overline{C}$ has the weakness to be relatively insensible to the magnitude of the temperature violations, as the factor $N^{-1}$ provides an average over the entire evaluation period. To overcome this issue the third metric, $\overline{\Delta}$,  estimates the mean temperature deviation per violating points i.e.,
 \begin{equation}
\overline{\Delta} = \frac{N}{\rm{N_\mathrm{out}}} \overline{C}
\end{equation}
where $\rm{N_\mathrm{out}}$ is the number of points lying outside the interval [19,24] \degree C. Finally,  the mean power electric consumption by the thermal systems are also reported to make sure that optimizers do not globally consume more than rule-based controllers:
\begin{equation}
\overline{P}_{\rm th}  = \frac{1}{N} \sum_{n=1}^{N} P_{\rm th}(t_n).
\end{equation}

All metrics are evaluated over two months in Barcelona, in February and March. The models were evaluated under the hypothesis of perfect weather predictions\footnote{Actual weather data was used in order to benchmark only the models and not errors from the predictions.}. The MPC controllers were compared against four rule-based controllers (abbreviated RB in Table \ref{table_mmpc}), that tried to keep the zones temperature close to the following constant setpoints: 19, 19.5 ,20 and 21 \degree C.  Results are reported on Table \ref{table_mmpc} and Figures \ref{mpc1} and \ref{mpc2}. Properties of rule-based controllers can be found in \ref{AA}, where rule-based commands for thermostats, tanks and heat pumps are given.

\begin{table}[h]
\vskip 0.15in
\begin{center}
\begin{small}
\begin{sc}
\begin{tabular}{llllll}
\toprule
\textbf{ Model}                                        & $\overline{P} $  (kW) & $\overline{P}_{ \rm th} $  (kW) &  $\overline{C}$  (\degree C) & $\overline{\Delta}$  (\degree C) \\
\midrule
RB 19 \degree C& 4.81  &4.46&1.35E-1&2.97E-1\\
RB 19.5 \degree C &5.13 &4.91&\textbf{3.48E-3}&\textbf{5.19E-2}\\
RB 20 \degree C& 5.41&5.32&0 &0\\
RB  21 \degree C  &6.01 &6.22& 0& 0\\
LSS-NL1  & \textbf{3.61} &\textbf{3.41}& 3.21E-2 &1.86E-1\\
LSS-NL2  & 3.94 &3.95& 1.93E-2 &2.40E-1\\
LSS-NL3  & 3.74 & 3.63&4.7E-3&2.42E-1 \\
LSS-NL4  &3.62 &3.56& 1.05E-1 &2.46E-1\\
LSS-NL5  &3.92&3.62 &5.11E-2 &2.23E-1\\
ENC-DEC1 & 4.03  &3.92&1.16E-2&1.50E-1\\
ENC-DEC2 &4.04&3.91&1.11E-2&1.57E-1\\
ENC-DEC3 & 4.32  &4.08&3.84E-3&9.37E-2\\
ENC-DEC4 &4.42&4.30&2.44E-2&1.76E-1\\
ENC-DEC5 &4.25&4.07&1.95E-2&1.61E-1\\
\bottomrule              
\end{tabular}
\end{sc}
\end{small}
\end{center}
\caption{ \small Key performance indicators for rule-based controllers and MPC on all model instances}
\label{table_mmpc}
\end{table}
\normalsize

From Table \ref{table_mmpc} and Figures \ref{mpc1} and \ref{mpc2},  the following key points can be observed:
\begin{itemize}
\item All optimizers, even the worst ones, consistently outperform the implemented rule-based controllers in terms of self-consumption and power consumption from thermal devices,

\begin{figure}[h]
  \centering
  \subfloat[Average power exchange vs. comfort violations \label{mpc1}]{\includegraphics[width=0.9\columnwidth]{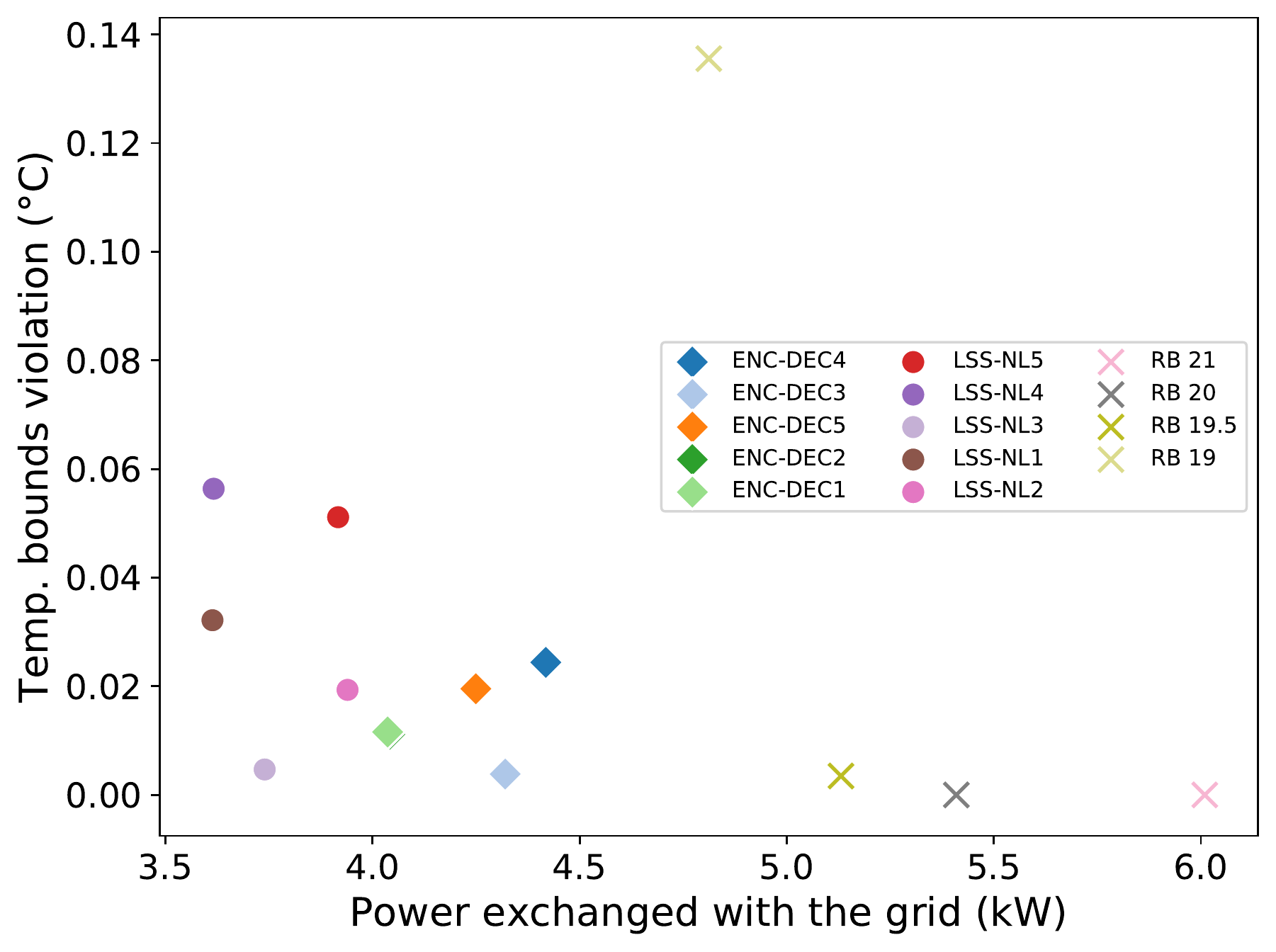}}
  \hfill
  \subfloat[ Violation rate vs. violation per deviation \label{mpc2}]{\includegraphics[width=0.8\columnwidth]{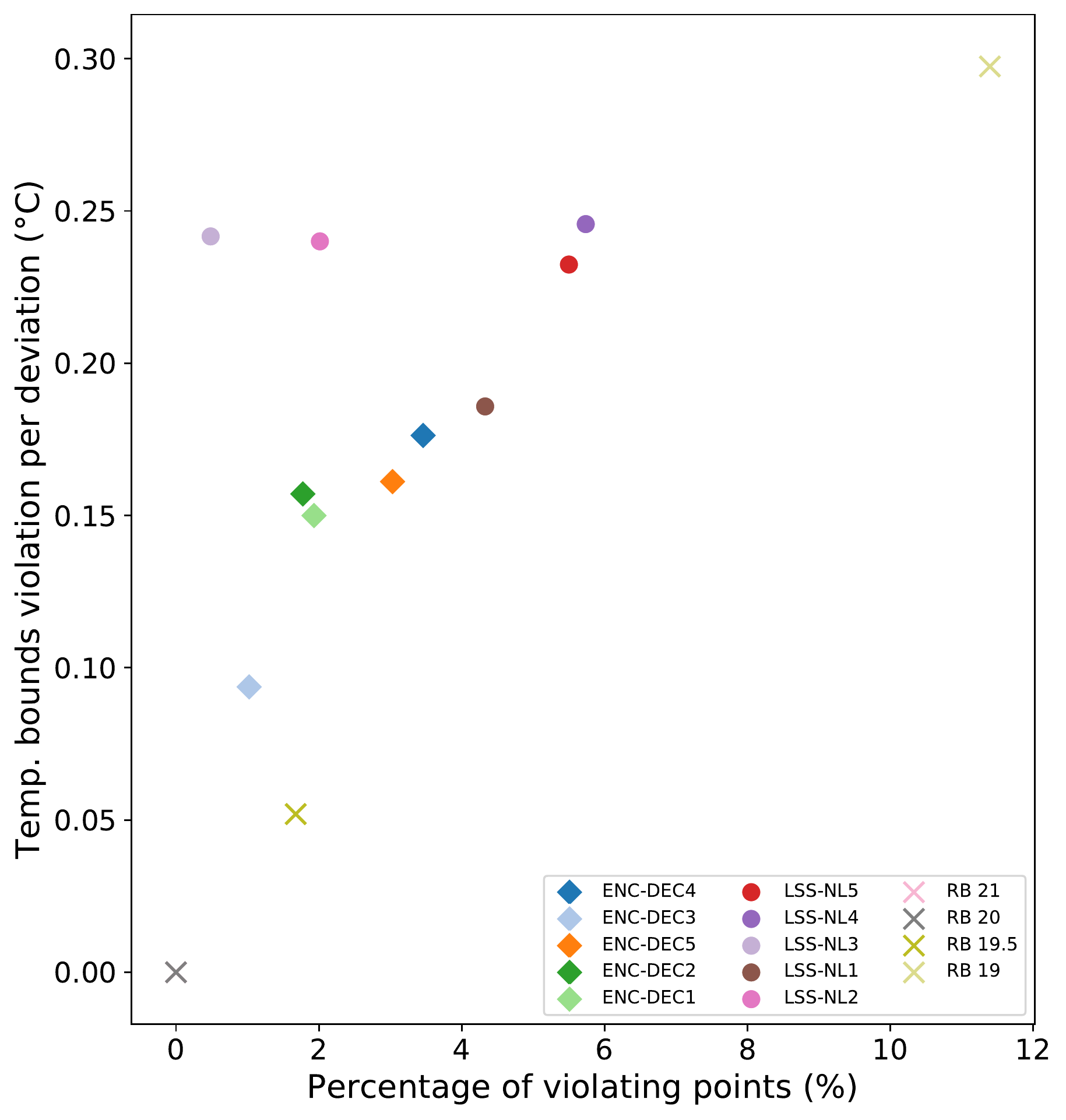}}
  \caption{ \small  (a) Mean self-consumed power  $\overline{P}$ and temperature deviation $\overline{C}$ for the model predictive controllers, as compared to rule-based controllers. ENC-DEC denotes encoder-decoder model instances, and LSS-NL linear state space model instances with non-linear power prediction; see Section \ref{sec2}. (b) Percentage of deviating points versus mean temperature violation per deviation}
\end{figure}

\normalsize
\item The Linear state space models show better results in power consumption metrics than the encoder-decoder models. However, as displayed clearly on Figure \ref{mpc2}, they come with the drawback of violating the constraints in a stronger way than the full non-linear models. On the trajectory displayed on Figure \ref{tmp_ex2}, the linear state space MPC is clearly violating the comfort rules at low temperature (light green curve), indicating that the model has not caught heating saturation when the heat pump temperature is too low. The non-linear MPC, however, only does a very minor violation. Scatter plots on Figure \ref{fig5b} clearly show the superiority of encoder-decoder models for comfort, as well as the low temperature excursions of the controllers with linear state space models.\\
\item To take advantage of the heat pump COP and the power generated by the photovoltaic panels, the two MPC architectures use a completely different strategy than typical heating curves used for rule-based controllers. At low outside temperature, both controllers lower the heat pump temperature to get a higher COP (Figure \ref{fig5a}).
\item Hours with higher ambient temperatures are correlated with higher irradiance, and thus a high probability of having excess of PV. So the controllers tend to increase the room setpoint temperatures as a strategy to preheat the buildings and reduce power exchange with the grid; see Figure \ref{fig5b} as well as  Figure \ref{tmp_ex1}.\\
\end{itemize}

\begin{figure}[h]
  \centering
  \subfloat[Heating curve \label{fig5a}]{\includegraphics[width=0.9\columnwidth]{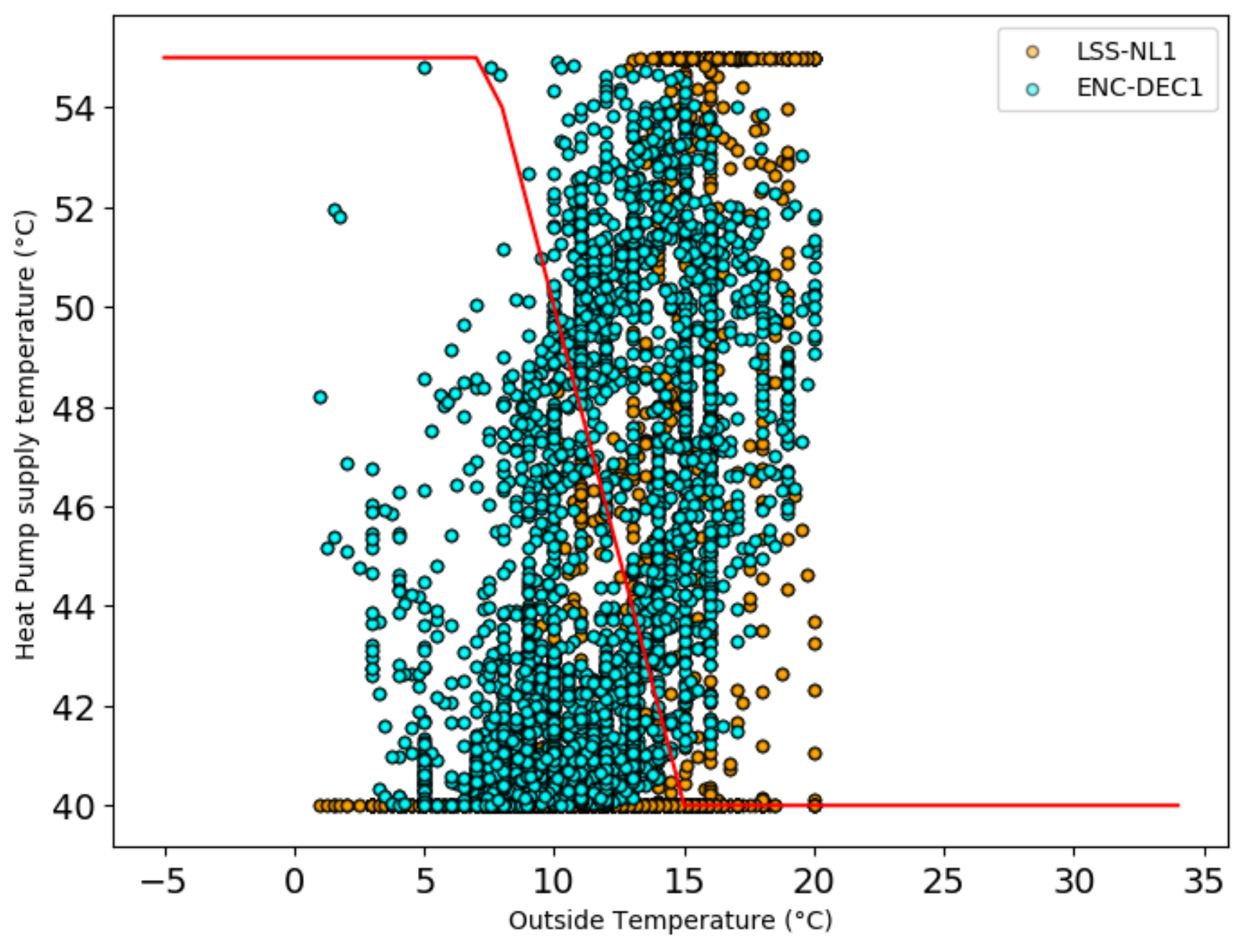}}
  \hfill
  \subfloat[Zone 1  room temperature measured \label{fig5b}]{\includegraphics[width=0.9\columnwidth]{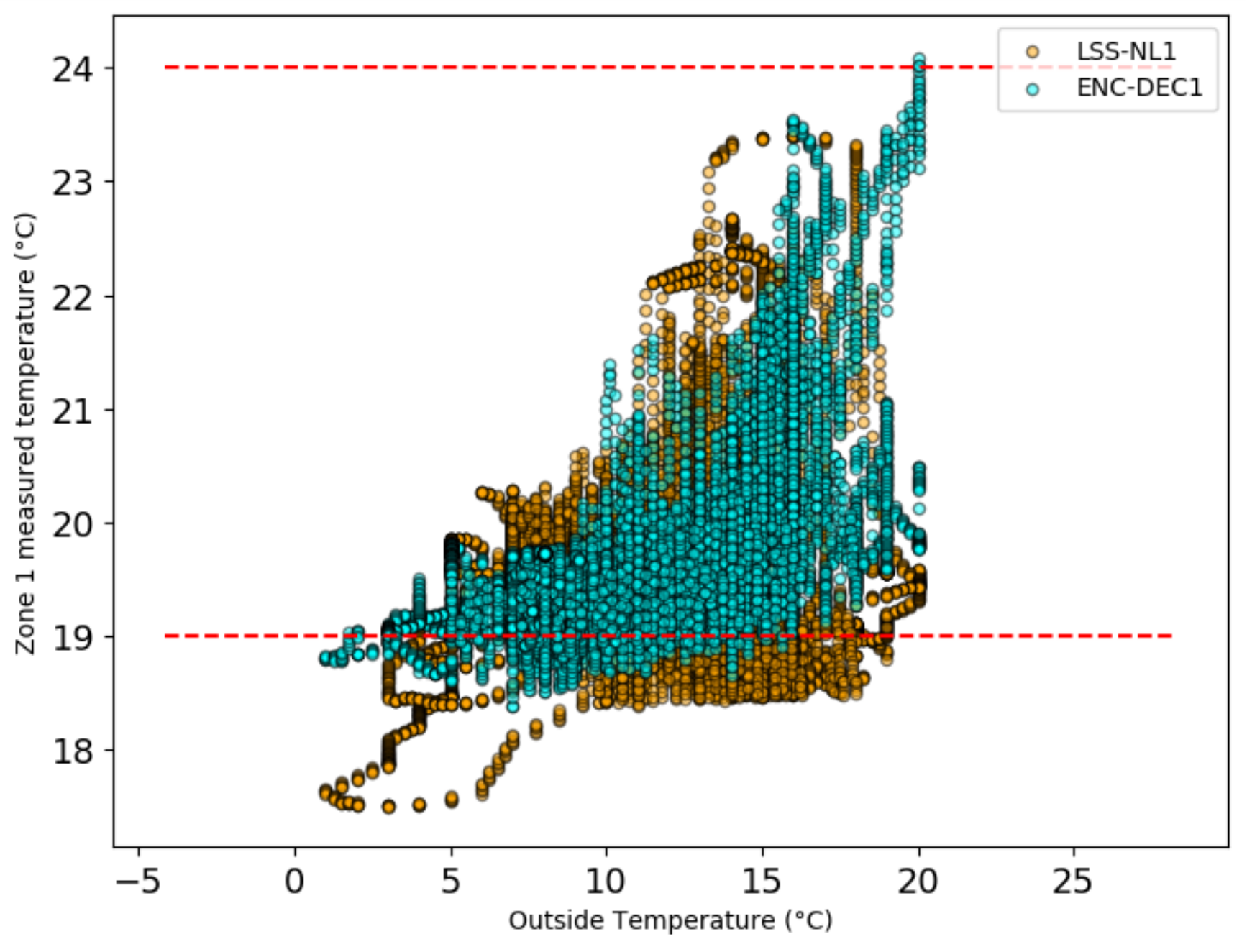}}
  \caption{ \small (a) Heating curve for the rule-based controllers (red), as compared to MPC Heat pump temperature supply scatter plots for two of the best performing models. (b) Zone 1  room temperature measured: Scatter plots for two of the best performing models. Linear state space controllers exhibit clear deviations at low outside temperatures.}
\end{figure}

\normalsize
These results show the difficulty of optimizing non-linear state space models with local optimization techniques. Local methods like SQP find it hard to escape a minimum and are quite sensitive to the initial trajectory solution. In the fully non-linear case, due to the high number of parameters and the non-convexity of the model, the objective hypersurface will present many local minima. On the contrary, for linear-state space models with smooth non-linear outputs (like the kernel regression used in this paper), fewer local minima are expected. This effect is well illustrated on the heat pump trajectories on Figure \ref{tmp_ex1}. The heat pump setpoints for the non-linear MPC (ENC-DEC1, in dark blue) slowly react to external conditions, and the optimizer takes time, at low temperature, to reach setpoints close to 40 \degree C. On the contrary, transitions from the linear state space model controller (LSS-NL1, in olive) are much faster, and setpoints quickly make the transition from high temperatures heat pump supply to the lower bound of 40 \degree C. Having a low supply temperature increases the COP of the heat pump, and therefore reduces energy consumption, especially at night. The fact that the SQP with the non-linear models takes more time to achieve the temperature transition lowers its energy performance.

\begin{figure}[h]
\begin{center}
\centerline{\includegraphics[width=\columnwidth]{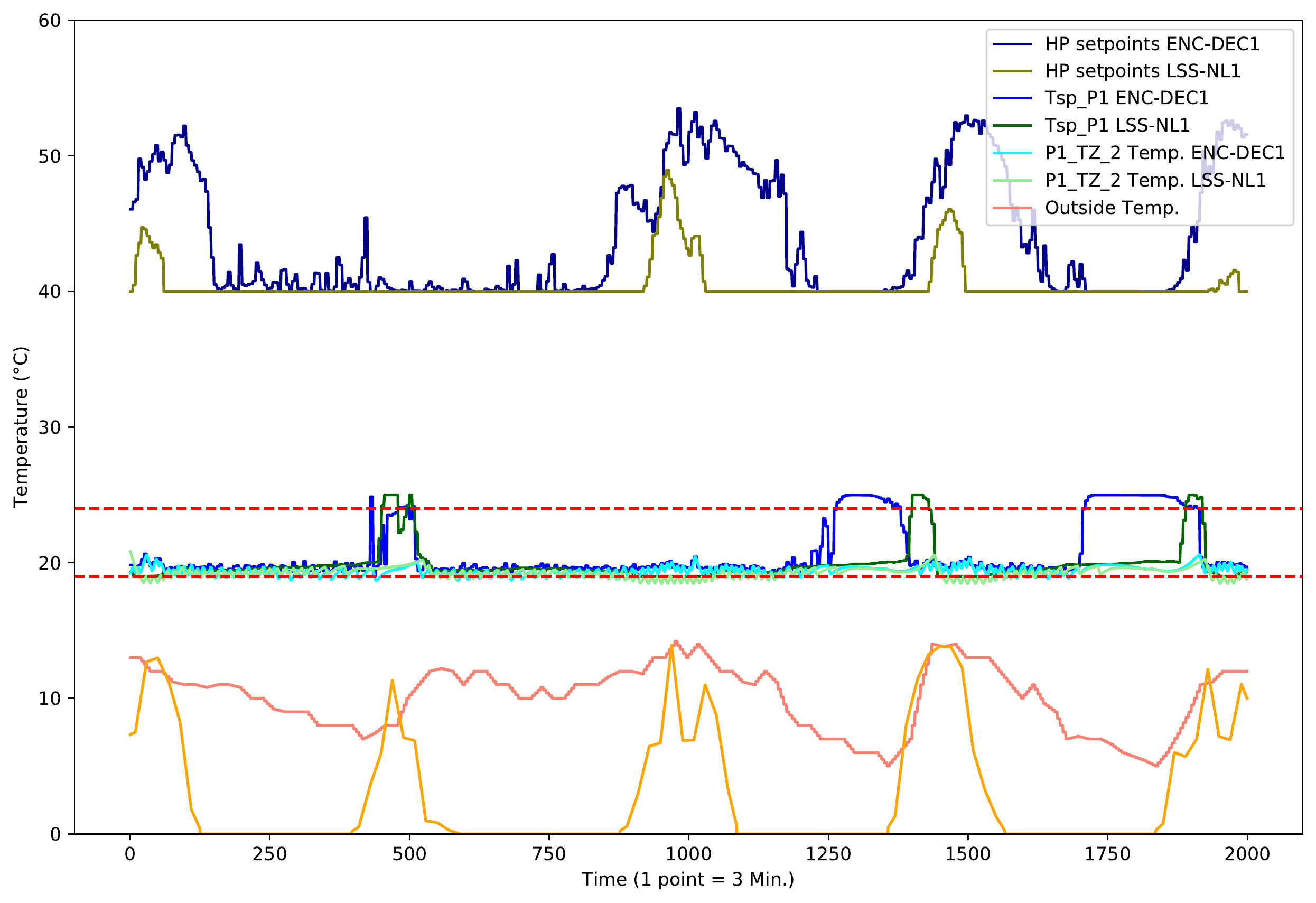}}
\caption{ \small Example of trajectories for  two of the best performing models of each architecture. Dashed red curves show the limit range of 19 and 24 \degree C. Irradiance (normalized) in orange.}
\label{tmp_ex1}
\end{center}
\end{figure}

\begin{figure}[h]
\begin{center}
\centerline{\includegraphics[width=\columnwidth]{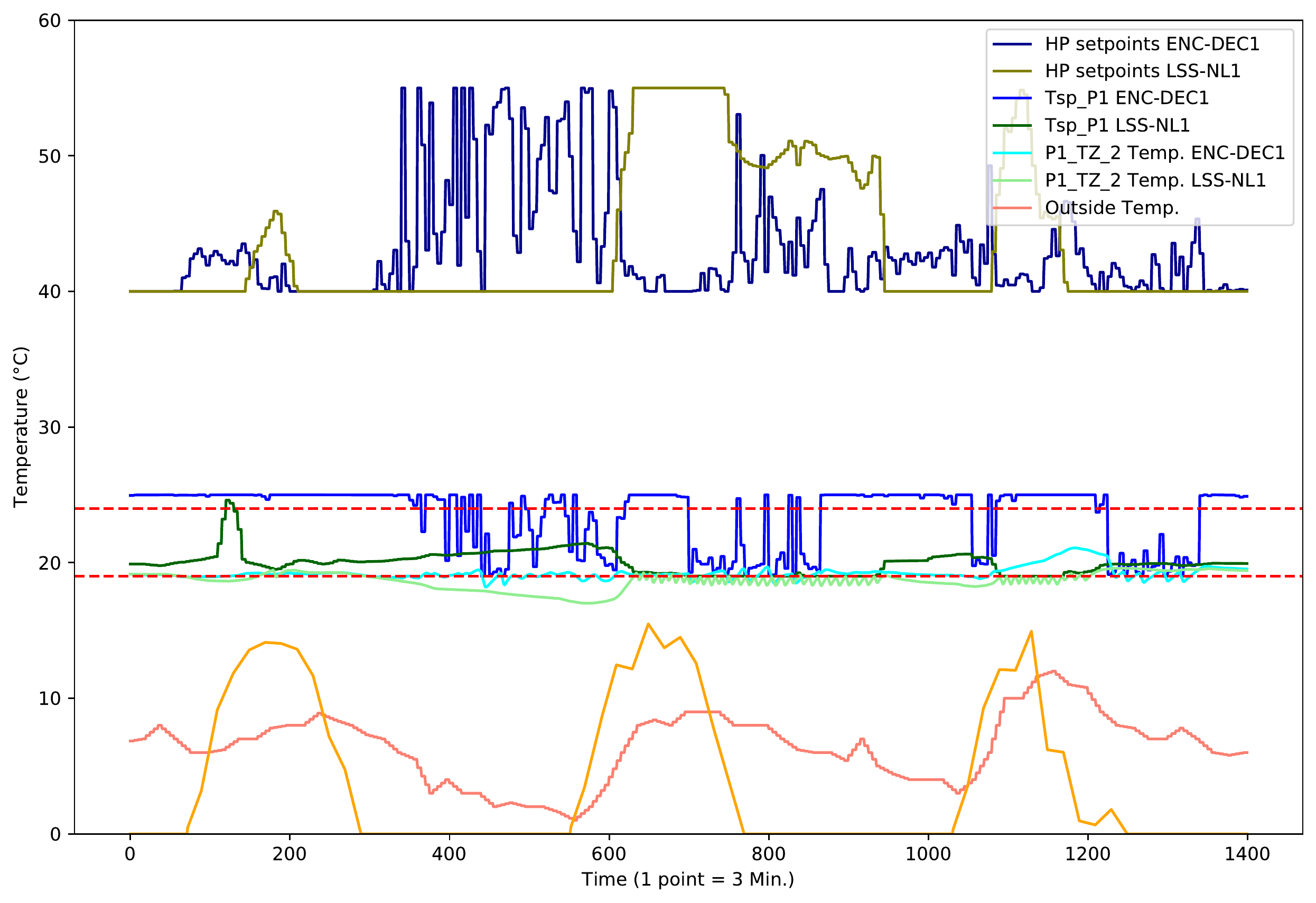}}
\caption{ \small Example of trajectories for  two of the best performing models of each architecture. Dashed red curves show the limit range of 19 and 24 \degree C. Irradiance (normalized) in orange.}
\label{tmp_ex2}
\end{center}
\end{figure}

\normalsize
\subsection{Execution time}
\label{excect}
The encoder-decoder architecture required about one hour for training (depending on the number of iterations) and the linear state-space model about one minute. In both cases this training time is negligible over the lifetime of the system in comparison with the computation time required to solve the optimal control problems.

For both architectures, the execution speed of the controllers was optimized. Evaluation and Jacobian's computations of the MPC with encoder-decoder models were carried out on a computing workstation with an  Nvidia RTX 2080 Ti Graphics Processing Unit (GPU). Jacobians in the SQP optimization were evaluated with Tensorflow 1.14 on the GPU for speed up, providing better time execution than numerical approximation with finite differences. Optimization coupled with the encoder-decoder models took around 4.5 minutes  per control step. Analytic computations of the Jacobians of the linear state-space model with non-linear outputs  were carried out using multiprocessing python package Numba \cite{lam2015numba} on a computing workstation with an Intel i9-9960X CPU at nominal frequency of 3.10GHz and 16 physical cores. Optimization coupled with the linear state-space models with non-linear outputs took around 1.3 minutes per control step. The two methods therefore require a computation time in the same order of magnitude at each control step but the linear state-space model results in an optimization more than three times as fast as the encoder-decoder architecture.

\section*{Conclusion}
\label{sec6}
 A comparison of two model architectures for MPC for building control has been presented: a linear state-space model combined with a non-linear regressor, and a fully non-linear architecture based on recurrent neural networks. Both architectures are capable of taking non-linearities into account. They were used within a local optimization procedure to minimize the power exchanged between a building and the grid without degrading comfort. 

Table \ref{table_weak_strong} presents a qualitative comparison of the two models used in the study. None comes on top of the other on all criteria. The second architecture based on RNNs is more data intensive but also more accurate. In general the higher representation capability of RNN architectures does not translate into strong enough control performance to justify the increased computational demand as compared to MPC based on well-designed linear state-space models. The exception to this statement would be cases were constraint violations have a disproportionate importance or for systems with more extreme non-linearities than the typical building used in this study.

\begin{table}[h]
\caption{Comparison summary of the architectures capabilities}
\begin{center}
\label{table_weak_strong}
\begin{small}
\begin{sc}
\begin{tabular}{lcc}
\toprule
 \textbf{Properties}     &  \textbf{LSS-NL} & \textbf{ENC-DEC}  \\
\midrule
Sample efficiency Sys. Id.  &  +   &  -  \\ 
Accuracy System Id.  &  -   &  +  \\ 
Objective minimization&  +  &  -  \\ 
Respect of constraints&  -  &  + \\ 
Computation time&  +  &  - \\
\bottomrule             
\end{tabular}
\end{sc}
\end{small}
\end{center}
\end{table} 

In future work the sample efficiency of the RNN architecture could be improved using pre-training  and transfer learning techniques for buildings that share the same inputs-outputs features. In cases where extreme non-linearities strengthen the value of this architecture research is needed to alleviate some of the drawbacks of the recurrent architectures outlined in Table \ref{table_weak_strong}. For instance, as suggested in \cite{maddalena2019neural}, it could be possible to reduce the computing costs by changing the encoder-decoder architecture to directly predict the quadratic form to be minimized in Eq. \eqref{quad_form}. 

Through this systematic comparison in the representative example of a building, we were able to demonstrate that, in their current form, RNN architectures do provide improved accuracy on the modeling of non-linear systems but have limited value in the optimal control of such systems. Well-designed linear state-space models with non-linear regressors are the solution of choice in most cases for model-predictive control.

\section*{Acknowledgment}
This project has received funding from the European Union’s Horizon 2020 research and innovation program under grant agreement n°731211, project SABINA.

\appendix
\section{Controller architectures}
\label{AA}
\subsection{Rule-based controllers}
\begin{figure}[th]
\begin{center}
\centerline{\includegraphics[scale=0.4]{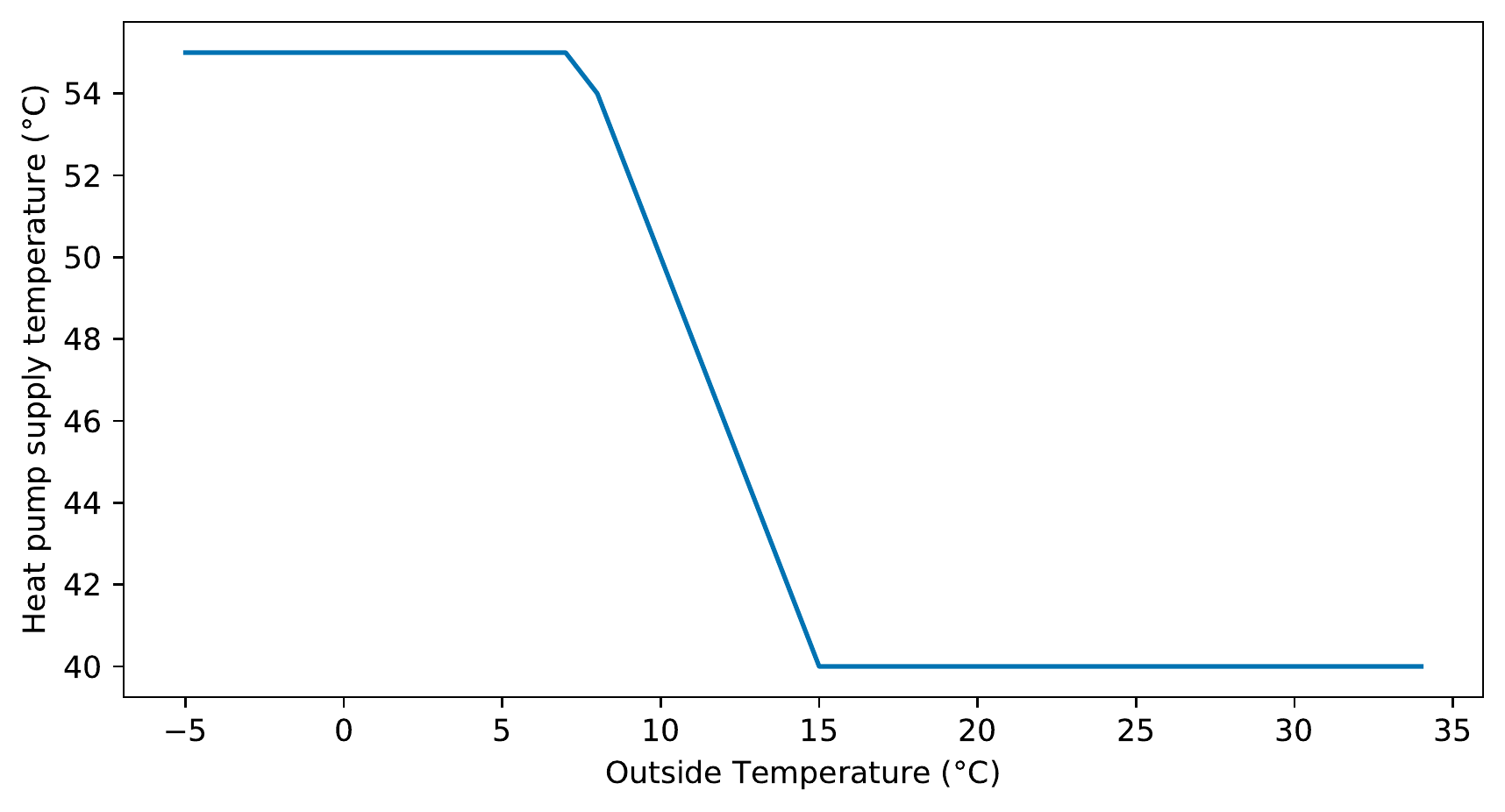}}
\caption{Heating curve for heat pump supply temperature}
\label{hc_ex}
\end{center}
\end{figure}

Rule-based controllers use fixed thermostat temperature setpoints for the rooms and a heating curve for the heat pump supply temperature. The heat pump temperature supply lies within  40 to 55 \degree C; see Figure \ref{hc_ex}. The heating strategy follows typical heating strategies in buildings, where supply temperature varies as a function of the outside temperature, providing more heat at lower temperature such as to ensure comfort bounds will not be violated. The heat pump supply and tank setpoints are the same for all models. The rule-based controller names indicate the fixed setpoints for the room thermostats: e.g., RB 19 is a rule-based controller with 19 \degree C temperature setpoint for the room thermostats. The following setpoints are used in the rule-based controllers:
\begin{table}[h]
\begin{center}
\label{rbcon}
\begin{small}
\begin{sc}
\begin{tabular}{lcc}
\toprule
Controllers    & Room sp. (\degree C) & Tank sp. (\degree C)\\
\midrule
RB 19& 19.0 & 40.0 \\ 
RB 19.5 & 19.5& 40.0 \\
RB 20  & 20.0& 40.0\\ 
RB 21 & 21.0& 40.0\\
\bottomrule             
\end{tabular}
\end{sc}
\end{small}
\end{center}
\caption{Rule-based controllers setpoints}
\end{table}

\quad

\subsection{Linear state-space models}
Linear state space models have been fitted using N4SID method. The models were fitted on data from the simulation obtained with random multisine excitations \cite{schoukens2019nonlinear}. The number of data points were varying between the models, as well as the excitations and building response.
\begin{table}[H]
\begin{center}
\label{sysid_lin}
\begin{small}
\begin{sc}
\begin{tabular}{lc}
\toprule
Models     &  Data points for training\\
\midrule
LSS-NL1 & 4000  \\ 
LSS-NL2  & 4000  \\
LSS-NL3  & 2000\\ 
LSS-NL4 & 2000\\
LSS-NL5  &750\\
\bottomrule             
\end{tabular}
\end{sc}
\end{small}
\end{center}
\caption{ \small Linear state-space models with non-linear outputs: Number of data points used for model identification, resolution of 15 minutes.}
\end{table}
The linear state-space models have 8 hidden degrees of freedom i.e., $x$ in Eq. \eqref{dyn_state_lin} belongs to  $\mathbb{R}^8$. Power energy consumption has been estimated with kernel ridge regression with rbf kernel with parameters $\gamma =0.1$ and $\alpha=1.0$ (see \cite{pedregosa2011scikit}). 

\subsection{Encoder-decoder models}
Encoder-decoder models use LSTM and multi-layer perceptrons as described on Figure \ref{nnfig}. All models have been identified with the same dataset but with distinct numbers of training epochs. Architecture parameters and number of training epoch per model instances are given in Tables . 
For training, following strategies were used:
\begin{itemize}
\item Batch size of 200,
\item Adam optimizer with default tensorflow  Adam parameters and learning rate of $1.10^{-4}$,
\item Decoding length randomly chosen at each iteration in the set $\{2,4,6,8,10,16,24,32,64,88,144\}$,
\item Gradient clipping with norm 1 (see \cite{zhang2019analysis} for an interesting analysis of gradient clipping and convergence speed up).
\end{itemize}

\begin{table}[ht]
\begin{center}
\begin{tabular}{lc}
\toprule
Layers     & hidden size\\
\midrule
LSTM encoder &512\\ 
LSTM decoder  &512 \\
layer1   &512\\ 
layer2  & 256\\
layer3   &128\\
\bottomrule             
\end{tabular}
\caption{  \small  Encoder-decoder sizing parameters}
\end{center}
\end{table}

\begin{table}
\begin{center}
\begin{tabular}{lc}
\toprule
Models     &  Training epochs\\
\midrule
ENC-DEC1 & $8.10^{4}$  \\ 
ENC-DEC2  & $8.10^{4}$  \\
ENC-DEC3 & $6.10^{4}$\\ 
ENC-DEC4 & $3.10^{4}$\\
ENC-DEC5  & $6.10^{4}$\\
\bottomrule             
\end{tabular}
\caption{  \small  Encoder-decoder training epochs per model instances (batch size of 200 sequences per epoch)}
\end{center}
\end{table}

\section{Error models summary}    % Each appendix must have a short title.
\label{error_app}
Models accuracy metrics on the evaluation set are given below for one hour ahead and one day ahead forecasts.
\small 
\begin{table}[h]
\label{sysid}
\begin{scriptsize}
\begin{sc}
\begin{tabular}{lcccc}
\toprule
Models     &  \textbf{MAE T} & \textbf{MAE P } & \textbf{sMRAE T} &\textbf{sMRAE  P}\\
\midrule
LSS-NL1 &0.97  &3.57 &0.047 &0.55\\ 
LSS-NL2  &0.73  &3.24&0.035 &0.51 \\
LSS-NL3  &0.77 & 3.52&0.037 &0.54\\ 
LSS-NL4 & 0.81 &3.43& 0.039&0.54\\
LSS-NL5  &0.71 &4.33& 0.034 & 0.63\\
ENC-DEC1 &0.29  & 2.25 & 0.014 &0.46\\
ENC-DEC2 & \textbf{0.26} & \textbf{1.91}& \textbf{0.012}&\textbf{0.36}\\
ENC-DEC3 &0.28 & 1.93 &0.013 &0.38\\
ENC-DEC4 &0.39  & 2.50 &0.019 &0.56\\
ENC-DEC5 &0.36  &2.04 & 0.017&0.39\\
\bottomrule             
\end{tabular}
\end{sc}
\end{scriptsize}
\caption{\small  Error metrics for sliding windows of one-day-ahead forecast, averaged over four months. Error metrics for sliding windows of one-hour-ahead, averaged over four months. Values corresponding to the best-performing model are in bold.}
\label{tableOneDayErrorMetrics}
\end{table}

\begin{table}[h!]
\label{sysid}
\begin{scriptsize}
\begin{sc}
\begin{tabular}{lcccc}
\toprule
Models     &  \textbf{MAE T} & \textbf{MAE P } & \textbf{sMRAE T} &\textbf{sMRAE  P}\\
\midrule
LSS-NL1 &0.64  &3.43 &0.030&0.54\\ 
LSS-NL2  &0.67  &3.24&0.031 &0.51 \\
LSS-NL3  &0.66 & 3.55&0.031 &0.55\\ 
LSS-NL4 & 0.77 &3.41& 0.037&0.54\\
LSS-NL5  &0.62 &4.32& 0.029 & 0.63\\
ENC-DEC1 &0.22  & 2.05 & 0.010 &0.40\\
ENC-DEC2 &\textbf{0.19} & \textbf{1.90}& \textbf{0.009}&\textbf{0.37}\\
ENC-DEC3 &0.19 & 1.93 &0.009 &0.38\\
ENC-DEC4 &0.24  & 2.50 &0.011 &0.54\\
ENC-DEC5 &0.20  &1.91 & 0.009&0.37\\
\bottomrule             
\end{tabular}
\end{sc}
\end{scriptsize}
\caption{  \small  Error metrics for sliding windows of one-hour-ahead, averaged over four months. Values corresponding to the best-performing model are in bold.}
\label{tableOneHourErrorMetrics}
\end{table}

%\clearpage
\small
\bibliographystyle{plain}
\bibliography{references}

\end{document}